
%
%
\input harvmac
\def\t{\theta}\def\tht{\theta}

\def\g{\gamma}
\let\p=\prime
\def\e{\epsilon}
\lref\Ki{P. Anderson, J. Phys. C3 (1970) 2436.}
\lref\Kii{K. Wilson, Rev. Mod. Phys. 47 (1975) 773.}
\lref\Kiii{P. Nozi\`eres, J. Low Temp. Phys. 17 (1974) 31; P. Nozi\`eres and
A. Blandin, J. Phys. (Paris) 41 (1980) 193.}
\lref\AFL{N. Andrei, K. Furuya and J. Lowenstein, Rev.
Mod. Phys. 55 (1983) 331.}
\lref\Ver{E. Verlinde, Nucl. Phys. B300 (1988) 360.}
\lref\CarVer{J. Cardy, Nucl. Phys. B324 (1989) 581.}
\lref\Gep{D. Gepner, ``Foundations of Rational Quantum Field Theory, I''
Caltech preprint CALT-68-1825.}
\lref\KRS{P. Kulish, N. Reshetikhin and E. Sklyanin, Lett. Math. Phys. 5
(1981) 393; E. Date, M. Jimbo, T. Miwa, M. Okado, Lett. Math. Phys. 12
(1986) 209.}
\lref\ALi{I. Affleck, Nucl. Phys. B336 (1990) 517;
I. Affleck and A. Ludwig, Nucl. Phys. B352 (1991) 849; Nucl. Phys. B360 (1991)
641.}
\lref\FT{L.D Faddeev, L.A. Takhtajan, Phys. Lett. A85 (1981) 375.}
\lref\ABI{D. Altsch\"uler, M. Bauer, C. Itzykson, Comm. Math. Phys. 132
(1990) 349.}
\lref\Witbos{E. Witten, Comm. Math. Phys. 92 (1984) 455.}
\lref\ALii{I. Affleck and A. Ludwig, Phys. Rev. Lett. 67 (1991) 161.}
\lref\RFR{N. Reshetikhin, J. Phys. A24 (1991) 3299; L. Faddeev and N.
Reshetikhin, Ann. Phys. 167 (1986) 227.}
\lref\RSOS{A.B. Zamolodchikov, Landau Institute preprint, September 1989.}
\lref\ZamoII{Al.B. Zamolodchikov, Nucl. Phys. B358 (1991) 497, 524.}
\nref\Zising{A.B. Zamolodchikov, Adv. Stud. Pure Math. 19 (1989) 1.}
\lref\rYY{C.N. Yang and C.P. Yang, J. Math. Phys. 10 (1969) 1115.}
\lref\AD{N. Andrei and C. Destri, Phys. Rev. Lett. 52 (1984) 364.}
\lref\Weig{P. Wiegmann, JETP Lett. 31 (1980) 379}
\lref\TW{A.M. Tsvelick and P.B. Wiegmann, Z. Phys. B54 (1985) 201; 
J. Stat. Phys. 38 (1985) 125; 
A.M. Tsvelick, J. Phys. C18 (1985) 159.}
\lref\FSZii{P. Fendley, H. Saleur and Al.B. Zamolodchikov, 
``Massless Flows II: the exact $S$-matrix approach'', BUHEP-93-9, USC-93-003,
LPM-93-08, hepth@xxx/9304051.}
\lref\ABF{G. Andrews, R. Baxter and J. Forrester, J. Stat. Phys. 35
(1984) 193.}
\lref\ZamZam{A.B. Zamolodchikov, Al.B. Zamolodchikov, Nucl. Phys. B379
(1992) 602.}
\lref\ZandZ{A.B. Zamolodchikov, Al.B. Zamolodchikov, Ann. Phys. 120 (1979)
253.}
\def\wgta#1#2#3#4{\hbox{\rlap{\lower.35cm\hbox{$#1$}}
\hskip.2cm\rlap{\raise.25cm\hbox{$#2$}}
\rlap{\vrule width1.3cm height.4pt}
\hskip.55cm\rlap{\lower.6cm\hbox{\vrule width.4pt height1.2cm}}
\hskip.15cm
\rlap{\raise.25cm\hbox{$#3$}}\hskip.25cm\lower.35cm\hbox{$#4$}\hskip.6cm}}
\def\wgtb#1#2#3#4{\hbox{\rlap{\raise.25cm\hbox{$#2$}}
\hskip.2cm\rlap{\lower.35cm\hbox{$#1$}}
\rlap{\vrule width1.3cm height.4pt}
\hskip.55cm\rlap{\lower.6cm\hbox{\vrule width.4pt height1.2cm}}
\hskip.15cm
\rlap{\lower.35cm\hbox{$#4$}}\hskip.25cm\raise.25cm\hbox{$#3$}\hskip.6cm}}
\def\wgtc#1#2#3#4{\hbox{\rlap{\vrule width1.3cm height.4pt}
\rlap{\lower.35cm\hbox{$#1$}}\rlap{\raise.25cm\hbox{$#2$}}
\hskip.38cm\rlap{\lower.6cm\hbox{\vrule width.4pt height1.2cm}}
\hskip.22cm
\rlap{\lower.35cm\hbox{$#4$}}\raise.25cm\hbox{$#3$}\hskip.3cm}}
\def\wgtd#1#2#3#4{\hbox{\rlap{\vrule width1.3cm height.4pt}
\rlap{\lower.35cm\hbox{$#1$}}\rlap{\raise.25cm\hbox{$#2$}}
\hskip.38cm\rlap{\lower.6cm\hbox{\vrule width1.5pt height1.2cm}}
\hskip.22cm
\rlap{\lower.35cm\hbox{$#4$}}\raise.25cm\hbox{$#3$}\hskip.3cm}}
\def\wgtdd#1#2#3#4{\hbox{\rlap{\raise.25cm\hbox{$#2$}}
\hskip.2cm\rlap{\lower.35cm\hbox{$#1$}}
\hskip.25cm\rlap{\vrule width1.9cm height.4pt}
\hskip1.0cm\rlap{\lower.6cm\hbox{\vrule width1.5pt height1.2cm}}
\hskip.25cm
\rlap{\raise.25cm\hbox{$#3$}}\hskip.25cm\lower.35cm\hbox{$#4$}\hskip.6cm}}
\Title{\vbox{\baselineskip12pt\hbox{BUHEP-93-10}}}
{\vbox{\centerline{Kinks in the Kondo problem}}}

\centerline{Paul Fendley}
\medskip\bigskip\centerline{Department of Physics, Boston University}
\centerline{590 Commonwealth Avenue, Boston, MA 02215, USA}
\centerline{\it fendley@ryan.bu.edu}
\bigskip

\vskip .3in 
We find the exact quasiparticle spectrum for the continuum Kondo problem of
$k$ species of electrons coupled to an impurity of spin $S$.  In this
description, the impurity becomes an immobile quasiparticle sitting on the
boundary. The particles are ``kinks'', which can be thought of as field
configurations interpolating between adjacent wells of a potential with $k+1$
degenerate minima.  For the overscreened case $k>2S$, the boundary has this
kink structure as well, which explains the non-integer number of boundary
states previously observed.  Using simple arguments along with the consistency
requirements of an integrable theory, we find the exact elastic $S$-matrix for
the quasiparticles scattering among themselves and off of the boundary. This
allows the calculation of the exact free energy, which agrees with the known
Bethe ansatz solution.

\bigskip
\Date{April 1993}
\vfill\eject

It is possible to solve integrable models directly in the continuum, without
recourse to a lattice Bethe ansatz.  First, one finds the spectrum by using
simple symmetry arguments extracted from conformal field theory, the
underlying lattice model, exact solutions of related models, etc.  The strict
requirements of an integrable theory allow these guesses to be made precise,
and exact quantities can then be derived.  This continuum approach is more
than just convenient: there are cases where it leads to results previously
unsuspected. The classic example is the critical Ising model in a magnetic
field, which is solvable only in the continuum \Zising.

The purpose of this paper is show how to apply these methods to the Kondo
problem and other integrable models with impurities. The idea is simple: one
starts with the quasiparticle description in the bulk, and then finds a
variety of ways of coupling the impurity while keeping the model integrable.
The only complication is in identifying what model has just been solved!

Here we find the exact quasiparticle spectrum in the general Kondo problem,
and show that these excitations are in fact kinks.  This gives a simple
qualitative picture and allows us to rederive the exact Bethe ansatz solution.
A nice feature is that everything is always finite: there is no Fermi sea to
fill because we study directly the excitations above the sea.  Moreover, we
give a simple explanation and derivation of the non-integer number of ground
states on the boundary. Here it follows from the restrictions on placing kinks
next to each other: with $q$ one-particle states, there are not necessarily
$q^N$ $N$-particle states.

The general Kondo problem is a three-dimensional non-relativistic problem,
consisting of $k$ species of massless free electrons antiferromagnetically
coupled to a single fixed impurity of spin $S$ by a term $\lambda \delta(x)
\sum_{i=1}^k {\bf S_j} \psi^\dagger_i {\bf\sigma^j} \psi_i$ in the Hamiltonian
(fermion spin indices are suppressed).  By looking at $s$-waves, we restrict
to the radial coordinate and this becomes a 1+1-dimensional problem where
fermions move on the half-line with the impurity at the boundary. Through a
variety of methods \refs{\Ki,\Kii,\Kiii}, it was found that there are two
critical points. At $\lambda=0$, there is a (high-temperature or UV) unstable
one where the impurity is decoupled. When this is perturbed, the model flows
to a (low-temperature or IR) strongly-coupled one where the electrons bond to
the impurity and ``screen'' its spin.  The Kondo temperature $T_K$ is the
scale at which the model crosses over from the region of one fixed point to
the other.  Numerous properties can be calculated exactly by using the Bethe
Ansatz \refs{\AFL,\Weig,\AD,\TW}, or by perturbed conformal field theory
\refs{\ALi,\ALii}. 

We must first understand the ``bulk'' properties of the model, which are
independent of the impurity coupling. Since the Kondo problem is
integrable\foot{This is obviously true since the Bethe ansatz solution exists.
In cases when such a solution is not known, one can use perturbed conformal
field theory to find the non-trivial conserved currents required for
integrability \Zising.}, we can find the exact quasiparticles and their exact
$S$-matrix in the bulk. The quasiparticles are the ``physical'' left-
and right-moving excitations on the half-line.  They are massless (i.e.\ have
no gap: $p=\pm E$) because without the impurity there is no scale in the
problem.  Because left-right scattering is trivial here, we can work on the
full line with only left movers: the particles with $x>0$ ($x<0$) are the
original left (right) movers and the impurity is at the ``boundary'' $x=0$.
We define the rapidity $\t_i$ of a left mover by $E=-p\equiv\exp(-\t_i)$.
Since the bulk problem is scale invariant the two-particle $S$-matrix can only
depend on the ratio of the two momenta, so we write this as $S_{LL}(\t)$,
where $\t\equiv\t_1-\t_2$.  Because the model is integrable, the individual
momenta of the particles do not change in a collision (complete elasticity)
and the $n$-particle $S$-matrix is the product of these two-body ones
(factorizability) \ZandZ.

The crucial observation is that in this continuum quasiparticle description,
the effect of the impurity is that of a single immobile particle sitting at
$x=0$.  We can derive the $S$-matrix for a left mover to scatter off of this,
because the integrability implies that this $S$-matrix element must satisfy
the same constraints as $S_{LL}$. The only dimensionless quantity is the ratio
of the particle's momentum to the Kondo temperature $T_K$; defining
$T_K\equiv\exp(-\t_K)$, the $S$-matrix can thus be written as $S_{BL}(\t)$
where here $\t\equiv \t_i - \t_K$. To understand what the ``boundary
particle'' actually is (i.e.\ what states it can have), we will look at the
qualitative behavior at the IR fixed point, but the exact solution extends all
the way up to the UV fixed point where the impurity decouples.

To find the quasiparticles in the bulk, we investigate the symmetries. Along
with the spin symmetry (which in the 1+1-dimensional reduction is an internal,
not a space-time, symmetry), we have a ``flavor'' symmetry interchanging the
$k$ species of electrons, as well a $U(1)$ charge symmetry.  These three
symmetries can be decoupled into the current algebras \refs{ \ALi,
\ABI}
\eqn\decomp{SU(2)_k \otimes SU(k)_2 \otimes U(1),}
where the subscript is the level of the affine Lie algebra. The technique of
non-abelian bosonization \Witbos\ means that a model with a $G_k$ current
algebra is equivalent to a sigma model where the fields take values
in the group $G$ and the WZW term is proportional to $k$. Thus our model in
the bulk can be described by the sum of three sigma models, one for each term
in \decomp.

Once the Kondo bulk piece is described in this manner, there is an important
simplification: the impurity (an $SU(2)$ spin) couples only to the $SU(2)_k$
sigma model \ALi. The other parts contribute only to bulk properties.  Thus
all we need is the quasiparticles of the $SU(2)_k$ sigma model, and these are
already known \refs{\RFR,\ZamZam,\FSZii}. The particles are massless, and form
doublets under the global $SU(2)$ symmetry. However, there is additional
structure: each particle is also a kink! Kinks occur in models with multiple
ground states. Classically, a kink $K_{ab}$ in one space dimension is a field
configuration which takes the value $a$ at spatial negative infinity and $b$
at positive infinity. In the quantum theory, this restricts multi-particle
configurations to be of the form $K_{ab}K_{bc}K_{cd}\dots$\ .  In our case,
the vacua $a$ run from $1$ to $k+1$, and allowed kinks interpolate between
adjacent vacua. This is pictorially described for $k$=3 in fig.\ 1. We label
the left-moving particle doublets by $(u_{a,a\pm 1}, d_{a,a\pm 1}) $.  The
$SU(2)$ symmetry rotates $u\leftrightarrow d$ without changing the vacuum
indices.

In the simplest case $k$=1, the only non-trivial structure is that of a
$(u,d)$ doublet; all the kinks do is go back and forth between the two wells.
The $SU(2)_1$ model is the continuum description of the spin 1/2 XXX spin
chain, so this statement is equivalent to saying that its spin waves have spin
1/2 \refs{\AFL,\FT}.

We can now determine what kinds of ``boundary particles'' there are. They
follow from the qualitative behavior at the infrared fixed point, which
depends crucially on the screening. In the underscreened case ($k<2S$), one
electron from each species binds to the impurity, effectively reducing the
impurity spin to $q\equiv S- (k/2)$.  In this case, the boundary particle can
be any member of a $2q+1$-dimensional $SU(2)$ multiplet. For example, for
$q=1/2$ the boundary is a ($u,d$) doublet under the $SU(2)$, just like the
bulk particles. In the exactly screened case ($k=2S$), the electrons
completely screen the impurity. Thus the boundary should not transform under
the $SU(2)$ and so it is a single particle. In neither of these cases is there
any reason to expect that the boundary has any kink structure.

The overscreened ($k>2S$) case is a little stranger. The impurity is still
completely screened and does not transform under the $SU(2)$, but there are
now ``leftover'' electrons. Since there is the flavor symmetry among the
electrons, the impurity must still couple to all of them. Therefore, if the
boundary is to have non-trivial structure, it must be a kink!  We have the
nice qualitative picture that in the underscreened case, the impurity couples
to the spin structure, while in the overscreened case it couples to the kinks.
First look at when there is one leftover electron ($k=2S+1$). Here we expect
that the boundary is a kink interpolating between adjacent vacua, just like
the bulk particles. The boundary, however, is not a ($u,d$) doublet because
spin has been screened out. In the general case with $p$ leftover
electrons ($k=2S+p$), we expect that the boundary is a ``multiple'' kink,
which can interpolate farther than just adjacent vacua. To make this precise
while keeping the model solvable, one uses a procedure called ``fusion''
\KRS. This is the kink version of what we did for the underscreened case.
There, to get a spin-$1$ boundary-particle, we multiplied two spin-$1/2$
representations and projected out the singlet. Here, one defines the boundary
``incidence'' matrix $I_p$, whose rows and columns correspond to the vacua;
$(I_p)^a_b=1$ if the vacua $a$ and $b$ are connected by a boundary kink and is
$0$ otherwise. The kinks in the bulk always have incidence matrix $I_1$, no
matter what $p$ is.  The $I_p$ follow from the analog of angular-momentum
multiplication:
\eqn\angmom{I_1 I_p = I_{p-1}+I_{p+1}}
where $I_0$ is the identity matrix. Thus for the $k=3$ case of fig.\ 1
$$I_1=\pmatrix{0&1&0&0\cr 1&0&1&0\cr 0&1&0&1\cr 0&0&1&0\cr}\qquad
I_2=\pmatrix{0&0&1&0\cr 0&1&0&1\cr 1&0&1&0\cr 0&1&0&0\cr}.$$
Thus for $k=3$ and $s=1/2$, the boundary spectrum consists of kink doublets
$13$, $31$, $24$, $42$, $22$ and $33$.

Knowing the spectrum on the boundary provides a simple way to understand the
ground-state degeneracy (i.e. the number of boundary states) at the critical
points \refs{\AD,\TW,\ALii}. This number is not necessarily an integer when
the volume of space is infinite. In the overscreened case it is not, a fact
which the boundary kinks explain nicely.  At the UV critical point, the
impurity is decoupled from the bulk, so the degeneracy is simply the number of
states of the impurity, which is $2S+1$.  For the underscreened and exactly
screened IR cases, the answer is equally simple: it is $2S-k +1$.  The
overscreened case presents an interesting problem: how many states is a
boundary kink? The question is easy to answer.  We represent the $j$th vacuum
by the vector $v_j=(0,0,...0,1,0,...)$ where the $1$ is in the $j$th place.
Multiplying this vector by the incidence matrix tells you what vacua are
allowed to be adjacent to it. Thus the $k$th entry of the vector $I_p I_1^N
v_j$ is the number of $N$-kink configurations with vacuum $j$ on one end and
$k$ on the other; the $I_p$ takes care of the fact that the boundary can
change the vacuum. The number of $N$-particle configurations with periodic
boundary conditions and $N$ large is simply $\lambda_p \lambda_1^N$, where
$\lambda_p$ is the largest eigenvalue of $I_p$.  (Since the bulk particles are
massless, in infinite volume there can be an arbitrary number of them even as
the temperature goes to zero, thus generically $N$ is large.)  The
contribution to the zero-temperature entropy coming from the boundary is thus
just $\log\lambda_p$; it is easy to show using \angmom\ that
\eqn\gsd{\lambda_p= {\sin{\pi (p+1)\over k+2}\over\sin{\pi \over k+2}}.}
This number is the largest eigenvalue of the structure-constant matrix
$n^p_{ab}$ \ALii, a fact which follows from a deep result in conformal field
theory \refs{\Ver,\CarVer}.  The matrix $n^p$ is related to the boundary
states in any conformal field theory \CarVer, which hints that kinks appear in
any conformal field theory, with $n^p$ taking the role of $I_p$; a similar
program has been proposed in \Gep.

Knowing the particle spectrum, the $S$-matrix is essentially fixed uniquely
by the constraints of factorizability, unitarity and crossing-symmetry
\foot{To be precise, it is unique up to the appearance of additional poles,
the so-called CDD ambiguity.  This ambiguity can be removed by calculating the
resulting central charge from the TBA: barring a strange cancellation, no
poles can appear without changing the answer.} \refs{\ZandZ,\ZamZam}.  For
$k$=1, the $S$-matrix has already been derived from the Bethe ansatz
\refs{\AFL,\FT}: the only massless two-particle $S$-matrix for a
doublet $(u,d)$ consistent with factorizability and $SU(2)$ symmetry is
\refs{\ZandZ,\ZamZam}
\eqn\LLsmat{\eqalign{S(u(\t_1)u(\t_2)
\rightarrow u(\t_2)u(\t_1))&= Z(\theta)\ (\t - i\pi)\cr
S(u(\t_1)d(\t_2)\rightarrow d(\t_2)u(\t_1))&= Z(\theta) \ \t\cr
S(u(\t_1)d(\t_2)\rightarrow u(\t_2)d(\t_1))&= Z(\theta)\  i\pi\cr}}
with a symmetry under $u\leftrightarrow d$.  Unitarity and crossing fix $Z$ to
be 
\eqn\ZLL{Z(\t)= {1\over \t -i\pi}\exp {i\over 2}
\int_{-\infty}^{\infty}{dt\over t} \sin t\theta
{e^{-\pi |t|/2}\over \cosh {t\pi\over 2}}.}

For general $k$, the simplest possibility (and the correct answer) is that the
scattering in the $SU(2)$ ($u,d$) labels is independent of the kink labels:
$$S_{LL} =S_{u,d}\otimes S_{\hbox{kink}}.$$
A two-kink configuration can be labeled by three vacua; a two-particle
$S$-matrix element can be labeled by four because only the middle vacuum can
change in a collision. The resulting massless kink $S$-matrix
\refs{\RSOS,\FSZii} is the RSOS solution of \ABF
\eqn\LLRSOS{\eqalign{&\wgtb {m}{m\pm1}m{m\mp1}  (\t) = \tilde Z(\theta)
\left({\beta_m\over\beta_{m+1}^{1/2}\beta_{m-1}^{1/2}}\right)^{i{\t\over\pi}}
\sinh\g(i\pi-\t)\cr
&\wgta {m\pm1}m{m\mp1}m(\t) =\tilde Z(\t)\left(
{\beta_{m+1}^{1/2}\beta_{m-1}^{1/2}\over \beta_m}\right)^{1+i{\t\over\pi}}
-\sinh\g\t \cr
&\wgta {m+1}m{m+1}m(\t) =
\tilde Z(\t)\left({\beta_{m+1}\over \beta_m}\right)^{i{\t\over\pi}}
{\beta_1\over \beta_m}\sinh\g(\t+im\pi)\cr
&\wgta{m-1}m{m-1}m(\t) = \tilde Z(\t)
\left({\beta_{m-1}\over \beta_{m}}\right)^{i{\t\over\pi}}
{\beta_1\over \beta_m}\sinh\g(im\pi-\t)\cr}}
where $\beta_m=\sinh(im\g\pi)$ and $\g=1/(k+2)$. The first element, for
example, describes the process $K_{m,m\mp 1} (\t_1) K_{m\mp1,m} (\t_2)
\rightarrow K_{m,m\pm 1}(\t_2) K_{m\pm 1,m}(\t_1)$.  
Remember that the allowed vacua run from $1$ to $k+1$, and adjacent vacua must
differ by $\pm 1$.  Unitarity and crossing restrict $\tilde Z(\t)$ to be
\eqn\Ztilde{\tilde Z(\t)= {1\over \sinh\g(\t -i\pi)}\exp {i\over 2}
\int_{-\infty}^{\infty}{dt\over t} \sin t\theta
{\sinh{(k+1)\pi t\over 2}\over \sinh{(k+2)\pi t\over 2} \cosh {t\pi\over 2}}.}
This $S$-matrix was effectively confirmed by calculating the correct bulk
central charge \RFR.

The boundary $S$-matrix $S_{BL}(\t)$ follows from the same constraints of
integrablility. For the overscreened case $p=1$, the boundary kink structure
is the same as the bulk, so $S_{BL}=S_{\hbox{kink}}$ as defined in \LLRSOS.
The fusion procedure then gives the $S$-matrix for arbitrary $p$ up to an
overall prefactor \KRS.  For $p=2$,
\eqn\fusion{\wgtd abcd (\t)\ \propto\ \sum_f\ \wgtc abgf (\t+{i\pi\over 2}) 
\quad\wgtc fgcd (\t-{i\pi\over 2})}
where the thick line denotes the boundary kink, and the $S$-matrix elements
on the right are those given in \LLRSOS. The construction ensures that the
result is independent of the choice of $g$.  One builds up the elements for
arbitrary $p$ by products of the form $S(\t+i(p-1)\pi/2)S(\t+i(p-3)\pi/2)\dots
S(\t-i(p-1)\pi/2)$. The prefactor $Z^{(p)}(\t)$ is
\eqn\Zp{Z^{(p)}\equiv\ \wgtdd{m+p}{m+p+1}{m+1}m= \exp {i\over 2}
\int_{-\infty}^{\infty}{dt\over t} \sin t\theta
{\sinh{(k+2-p)\pi t\over 2}\over \sinh{(k+2)\pi t\over 2} \cosh {t\pi\over
2}},}
where we suppressed the $\beta_m$. 

The underscreened case proceeds in the same manner. When $k=2S-1$, the
boundary particle is an $SU(2)$ doublet like the bulk ones, so
$S_{BL}=S_{u,d}$ as defined in $\LLsmat$. For general underscreening with
$q\equiv S-(k/2)$, we use the spin analog of \fusion. The analog of \Zp\ is
that the $S$-matrix element for scattering a $u$ bulk particle with the
highest member ($S_z=q$) of the boundary multiplet is given by $Z^{(2q)}(\t)$
from \Zp\ with $k\rightarrow\infty$.

For the exactly screened case, the answer is not as obvious because the
boundary particle is neither a kink nor does it have any $SU(2)$ structure.
The simplest non-trivial solution of the consistency requirements is 
\eqn\SRL{S_{BL}=-i\tanh\left({\t\over 2}-{i\pi\over 4}\right).}
This result also has some simple analogues. Because of the lack
of structure of the boundary in the IR the irrelevant operator by which one
perturbs is simply the left-moving energy-momentum tensor $T_L$ \ALi. In
the similar flows from the tricritical Ising model to the Ising model
\ZamoII\ and from the $SU(2)_1$ principal chiral model into the WZW
model \ZamZam, the irrelevant perturbing operator is $T_L T_R$. Both of these
cases have a $LR$ $S$-matrix of \SRL\ (without the $i$), so it is not
surprising this is true here as well.

With the exact $S$-matrices, we can calculate the exact free energy using the
thermodynamic Bethe ansatz (TBA) \ZamoII, where one finds the allowed momenta
for the particles on a circle of circumference $l$, and then uses this
constraint to minimize and hence derive the free energy. It is similar to
ordinary Bethe ansatz thermodynamics \rYY, but here we work with the
``physical'' quasiparticles instead of the ``bare'' lattice electrons
previously used \refs{\AD,\TW}. In this approach there are no infinities.  We
quantize a momentum $p_i$ by demanding periodicity of the wavefunction when
the particle is ``brought around the world'':
\eqn\quant{e^{ip_i l}\Lambda(\tht_i| \t_K; \tht_1, \tht_2,\dots,\tht_N)=1,}
where $\Lambda$ is the eigenvalue for scattering one particle through an
ensemble of all the others and the impurity. When the $S$-matrix is diagonal,
$\Lambda=\prod_j S(\t_i-\t_j)$; in our non-diagonal case one must use some of
the formal tools of the Bethe ansatz to find it. For both parts of our
tensor-product $S$-matrix this has already been done
\refs{\ZamoII,\ZamZam,\FSZii}; we only need to add the effect of the
boundary particle.  This does not affect the bulk part of course, which is
still expressed in terms of a set of ``pseudo-energy'' functions $\e_j(\tht)$
obeying the integral equations
\eqn\TBA{
\e _j(\t )= \delta_{jk} e^{-\t} -\int {d\t ^{\p}\over 2\pi} {1\over
\cosh(\t-\t^{\p})}\left(\ln(1+e^{-\e _{j-1}(\t ^{\p})})
+\ln(1+e^{-\e _{j+1}(\t ^{\p})})\right)}
for $j=1\dots\infty$, and $\epsilon_0\equiv \infty$. Those from $1$ to $k-1$
arise from ``diagonalizing'' the kink part of the $S$-matrix, while those from
$k+1$ on up come from the $u,d$ part. The bulk free energy depends on
temperature trivially because there is no other scale in the problem;
including all pieces from \decomp\ it turns out to be $f_{bulk}=-k\pi T^2 l/
12$. The presence of the boundary particle in \quant\ merely adds an extra
term to the level densities, which then adds an $l$-independent piece to the
free energy. One finds that
\eqn\impur{f_{imp}= {T\over 2 \pi}\int d\t {1\over \cosh(\t-\ln(T/T_K))} 
\ln (1+ e^{-\e_{2S}}). }
I hope to present details and applications to similar problems in the future.

\bigskip\smallskip
I would like to thank Mark Goulian, Ken Intriligator and Hubert Saleur for
useful conversations. This work was supported in part by DOE grant
DEAC02-89ER-40509. 
\listrefs
\end